\documentclass[a4paper]{article}

\usepackage{INTERSPEECH2019}

\title{Unsupervised Representation Learning with Future Observation Prediction for Speech Emotion Recognition}
\name{Zheng Lian$^{1,3}$, Jianhua Tao$^{1,2,3}$, Bin Liu$^{1}$, Jian Huang$^{1,3}$}
\address{
$^{1}$National Laboratory of Pattern Recognition, CASIA, Beijing, China\\
$^{2}$CAS Center for Excellence in Brain Science and Intelligence Technology, Beijing, China\\
$^{3}$School of Artificial Intelligence, University of Chinese Academy of Sciences, Beijing, China}
\email{\{zheng.lian, jhtao, liubin, jian.huang\}@nlpr.ia.ac.cn}

\begin{document}

\maketitle
\begin{abstract}
Prior works on speech emotion recognition utilize various unsupervised learning approaches to deal with low-resource samples.
However, these methods pay less attention to modeling the long-term dynamic dependency, which is important for speech emotion recognition.
To deal with this problem, this paper combines the unsupervised representation learning strategy -- Future Observation Prediction (FOP), with transfer learning approaches (such as Fine-tuning and Hypercolumns). 
To verify the effectiveness of the proposed method, we conduct experiments on the IEMOCAP database. Experimental results demonstrate that our method is superior to currently advanced unsupervised learning strategies.

\end{abstract}
\noindent\textbf{Index Terms}: speech emotion recognition, unsupervised learning, Future Observation Prediction, transfer learning

\section{Introduction}

Obtaining large amounts of realistic data is currently challenging and expensive in speech emotion recognition \cite{schuller2011recognising}. Compared with automatic speech recognition which has thousands of hours of transcribed speech, databases annotated in emotional categories are still scarce. For example, the eNTERFACE \cite{martin2006enterface} and EMODB \cite{burkhardt2005database} databases contain less than 1000 samples. Although the IEMOCAP \cite{busso2008iemocap} and FAU-AIBO \cite{batliner2004you} databases contain 10039 and 13348 samples respectively, the distribution of emotional categories is extremely unbalanced.

Prior works on speech emotion recognition utilize various methods to deal with low-resource training samples, including unsupervised representation learning \cite{deng2014introducing,eskimez2018unsupervised} and transfer learning \cite{pan2010survey, ouyang2017audio}. Unsupervised representation learning takes full advantage of the information from unlabeled data. It does not utilize the label information but aims to learn robust representations that can capture intrinsic structures of the data. While transfer learning makes use of additional labeled data from a different but related task. Its main idea is to share the ``knowledge'' from the source task to the target task.

Most of existing unsupervised learning approaches provide salient representations, leading to notable improvement for speech emotion recognition \cite{deng2014introducing, eskimez2018unsupervised,xia2013using, ghosh2016representation}. A mainstream approach of unsupervised learning is to train autoencoders (AE) \cite{poultney2007efficient,bengio2007greedy}, including denoising autoencoders (DAE) \cite{vincent2010stacked}, variational autoencoders (VAE) \cite{kingma2013auto} and autoencoders with ladder structures \cite{rasmus2015semi}.
These methods \cite{poultney2007efficient,bengio2007greedy} take the whole input into account, aiming to learn intermediate feature representations that can reconstruct the input. However, they pay less attention to modeling the long-term dynamic dependency, which is important for speech emotion recognition \cite{huang2016attention, kim2017towards}.

To deal with this problem, this paper focuses on another popular unsupervised learning strategy, whose main idea is to predict future, missing or contextual information \cite{oord2018representation,mikolov2013efficient,zhang2016colorful,doersch2015unsupervised}. For convenience, it is marked as Future Observation Prediction (FOP). As FOP needs to infer the global information in the prediction process, capturing long-range dynamic information is also important \cite{oord2018representation,mikolov2013efficient}. The main intuition behind FOP is to learn the representations that encode the underlying shared information between different parts of the signal. In time series, FOP uses next step prediction to exploit the local smoothness of the signal, and discards low-level information and noise that are local. When predicting further in the future, the amount of shared information become much lower, and the FOP model is able to infer more global structures \cite{oord2018representation}. Recent works have successfully used the idea behind FOP to learn robust and generic representations in audio, textual and visual domains \cite{oord2018representation,mikolov2013efficient,zhang2016colorful,doersch2015unsupervised}. Mikolov et al. \cite{mikolov2013efficient} learned word representations by predicting neighboring words. Zhang et al. \cite{zhang2016colorful} learned image representations by predicting color for grey-scale images. Oord et al. \cite{oord2018representation} proposed a general framework to learn effective representations for multiple domains.

To capture long-term dynamic dependencies, we propose to use the self-attention mechanism \cite {vaswani2017attention} for FOP. As the self-attention mechanism can provide an opportunity for injecting global context of the whole sequence into each input frame, it can attend to longer sequences than many typical RNN-based models \cite{vaswani2017attention,liu2018generating}.

FOP can learn discriminative representations from the inputs. To distill the ``knowledge'' from FOP to emotion recognition, two transfer learning approaches are utilized, including Fine-tuning \cite{ouyang2017audio,thrun2012learning} and Hypercolumns \cite{Hariharan2015Hypercolumns}. 
Fine-tuning \cite{ouyang2017audio,thrun2012learning} is an approach that fine-tunes either the last or several of the last layers in the pre-trained model and leaves the remaining layers unchanged.
While Hypercolumns \cite{Hariharan2015Hypercolumns} concatenate embeddings at different layers in a pre-trained model. The target task can benefit from Hypercolumns as they connect low-level features with high-level semantics.

This paper combines the unsupervised representation learning strategy -- FOP, with transfer learning approaches (such as Fine-tuning and Hypercolumns) to improve the performance of speech emotion recognition. The main contributions of this paper lie in three aspects: 1) to deal with long-range temporal dependencies, we propose to utilize the self-attention mechanism for FOP; 2) to share the ``knowledge'' from FOP to speech emotion recognition, two transfer learning strategies are evaluated; 3) our proposed method is superior to other currently advanced unsupervised learning strategies. To the best of our knowledge, it is the first time that FOP is utilized as the unsupervised learning strategy to improve the performance of speech emotion recognition.

\section{Proposed Method}
Our training procedure consists of two stages. In the first stage, we train a model for FOP to learn discriminative representations from the inputs, which is marked as the \textbf{FOP model}. In the second stage, to share the ``knowledge'' from FOP to speech emotion recognition, we take advantage of transfer learning approaches including Fine-tuning (in Fig. 1(a)) and Hypercolumns (in Fig. 1(b)).

\begin{figure}[h]
	\centering
	\includegraphics[width=\linewidth]{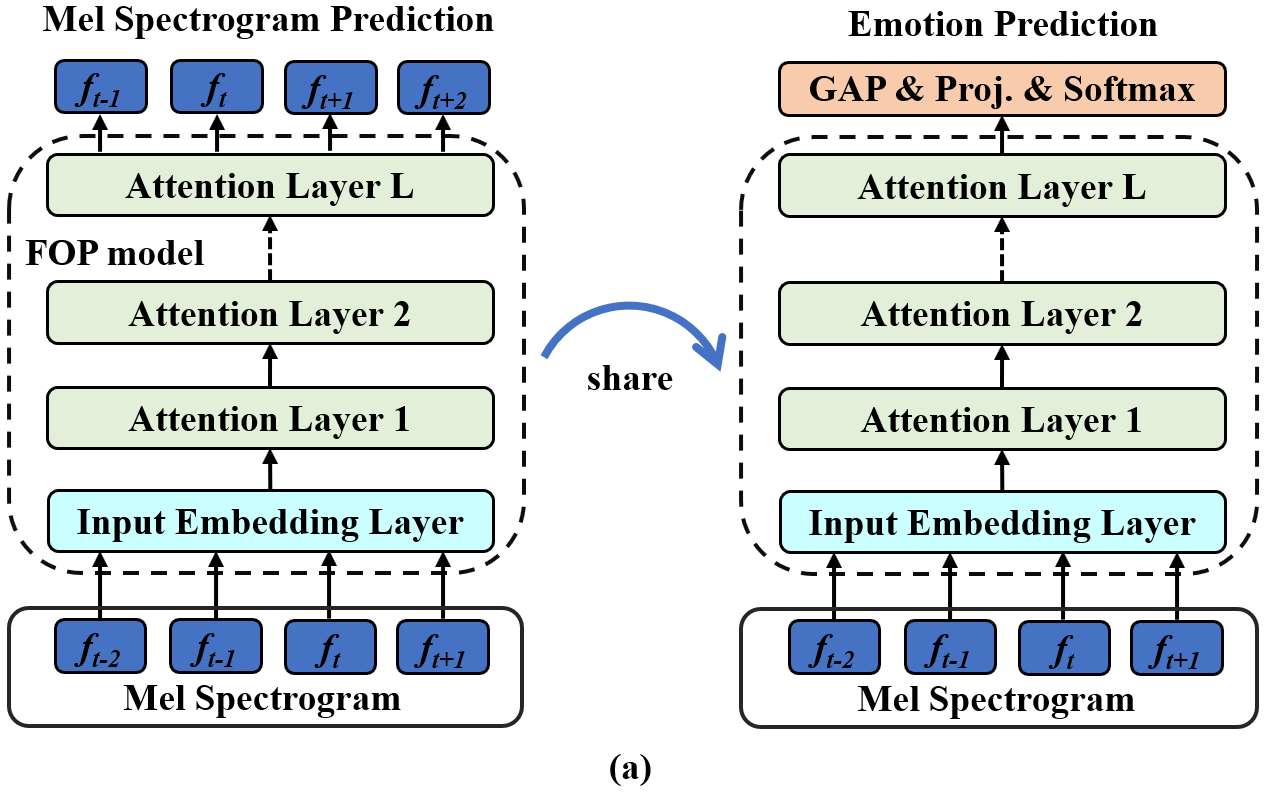}
	\includegraphics[width=\linewidth]{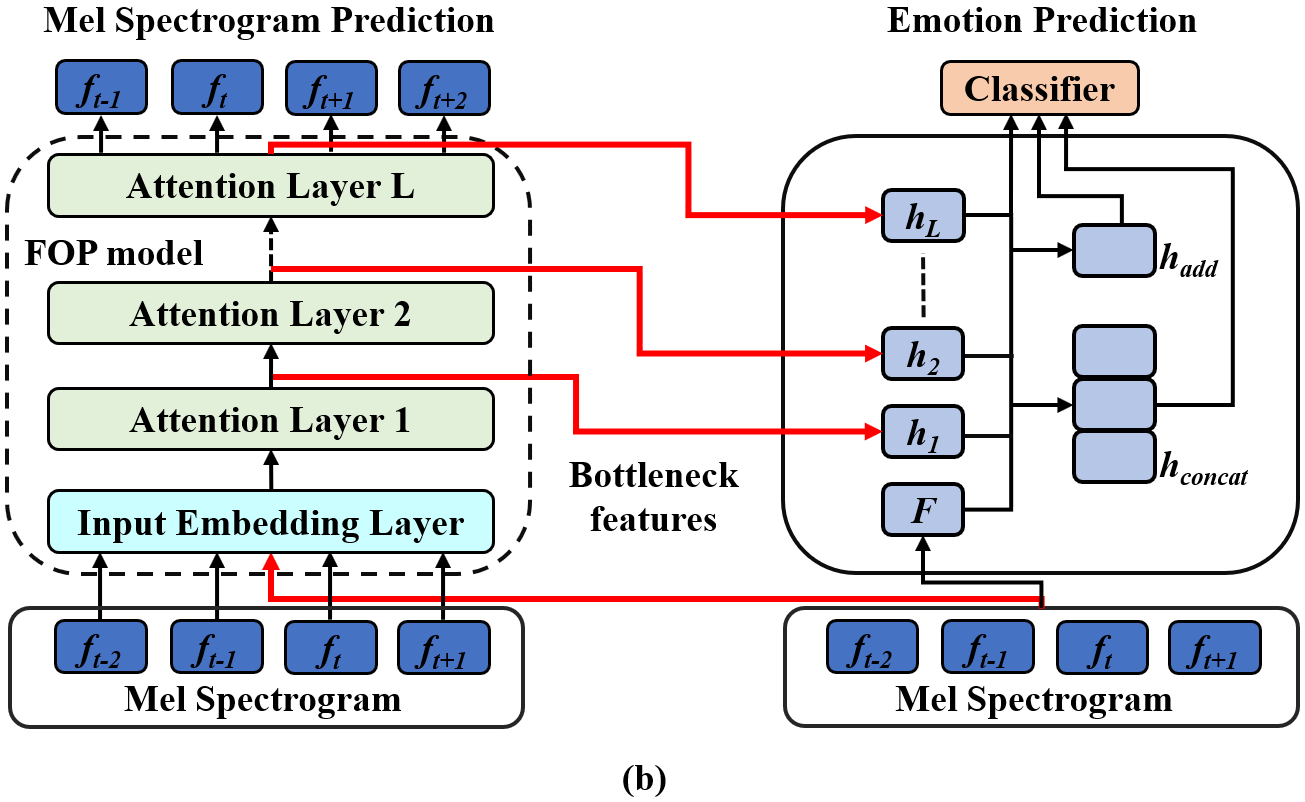}
	\caption{Overall structure of the proposed framework. (a) FOP combined with Fine-tuning: Through unsupervised pre-training, the FOP model can learn discriminative representations from the inputs. Then the output layer of the pre-trained FOP model is replaced by the combination of a global average pooling (GAP) layer and an emotion-specific projection layer. Finally, the whole model is fine-tuned for speech emotion recognition. (b) FOP combined with Hypercolumns: Bottleneck features are extracted from the pre-trained FOP model, followed with additional classifiers for speech emotion recognition.}
	\label{fig:system}
\end{figure}

\subsection{Unsupervised pre-training}
FOP can learn discriminative representations from the inputs.
This paper utilizes a generative model operating on audio features for FOP.
As depicted in Fig. 1, given an 80-dimensional mel-scale spectrogram $F=[f_{1},f_{2},...,f_{T}]$, where $T$ is the number of frames and $f_{t}$ is the features at timestep $t$. The joint probability of a feature sequence $F$ is factorized as a product of conditional probabilities, which is computed as follows:
\begin{equation}
	P(F)=\prod\limits_{t=1}^T P(f_{t}|f_{1},f_{2},...,f_{t-1};\theta)
\end{equation}
where the conditional probability $P$ is modeled using neural networks with parameters $\theta$. Each $f_{t}$ is therefore conditioned on the features at all previous timesteps $f_{1},f_{2},...,f_{t-1}$.

Capturing long-term dynamic dependencies is important for FOP. Several architectures are evaluated, including convolutional neural networks (CNNs) \cite{lecun1995convolutional}, long-short term memory (LSTM) \cite{hochreiter1997long} and the masked multi-head self-attention mechanism \cite{vaswani2017attention}. We find that the last one is more suitable for FOP. As the self-attention mechanism can inject global context of the whole sequence into each input frame, it can attend to longer sequences than other two architectures. Therefore, the conditional probability distribution is modeled by a stack of masked multi-head self-attention mechanisms \cite{vaswani2017attention}.

As depicted in Fig. 1, the FOP model has an input embedding layer and $L$ identical attention layers. The input embedding layer injects position information into the inputs. The attention layer captures long-term dynamic dependencies. We describe these components below.

\begin{figure}[h]
	\centering
	\includegraphics[width=0.9\linewidth]{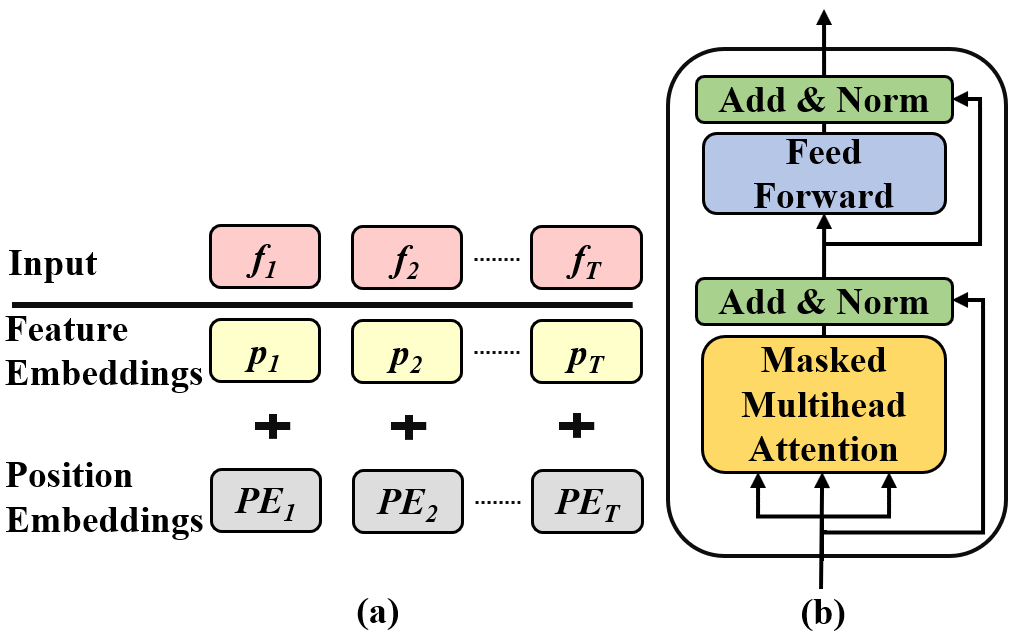}
	\caption{Two components in the FOP model. (a) Input embedding layer: The input embeddings are the sum of corresponding feature embeddings and position embeddings. (b) Attention layer: The attention layer has two sub-modules: a masked multi-head self-attention layer \cite{vaswani2017attention} and a feed forward layer.}
	\label{fig:self-attention}
\end{figure}

\subsubsection{Input embedding layer}
We first describe the input embedding layer, illustrated in Fig. 2(a). The FOP model relies on the self-attention mechanism, which contains no recurrence and no convolution networks. If we shuffle the inputs $F=[f_{1},f_{2},...,f_{T}]$, we will get the same output \cite{vaswani2017attention}. To take the order of the sequence into consideration, we inject the position information into $F$.

This paper utilizes position embeddings the same with that in \cite{vaswani2017attention}. These embeddings are computed as follows:
\begin{equation}
	PE_{(pos,2i)}=sin(pos/10000^{2i/d^{model}})
\end{equation}
\begin{equation}
	PE_{(pos,2i+1)}=cos(pos/10000^{2i/d^{model}})
\end{equation}
where $pos$ is the time step index, $i$ is the dimension and $d^{model}$ is the vector dimension of each frame.

The input embeddings $z_{t}$ are the sum of corresponding feature embeddings $p_{t}$ and position embeddings $PE_{t}$.
\begin{equation}
	p_{t}=W_{e}f_{t}
\end{equation}
\begin{equation}
	z_{t}=p_{t}+PE_{t}
\end{equation}
where $W_{e}$ is the feature embedding matrix and $f_{t}$ is the inputs at timestep $t$.

\subsubsection{Attention layer}

The input embeddings for all timesteps $h_{0}=[z_{1},z_{2},...,z_{T}]$ are passed into a stack of $L$ identical attention layers. The outputs of each attention layer are marked as $h_{l}, l \in [1,L]$. As depicted in Fig. 2(b), each attention layer has two sub-modules. The first is a masked multi-head self-attention mechanism \cite{vaswani2017attention}, and the second is a feed forward layer. Then we employ a residual connection \cite{he2016deep} around two sub-modules, followed with layer normalization \cite{ba2016layer}. Finally, the outputs of the last layer $h_{L}$ is utilized to predict the next frame's features.

The main ingredient of the attention layer is the masked self-attention mechanism \cite{vaswani2017attention}. With the help of this mechanism, we make sure that the prediction $P(f_{t}|f_{1},f_{2},...,f_{t-1})$ at timestep $t$ cannot depend on any of the future timesteps features $f_{t},f_{t+1},...,f_{T}$.

\subsection{Supervised transfer learning}
To take advantage of discriminative representations that are learned from FOP, two transfer learning approaches are evaluated, including Fine-tuning and Hypercolumns.

\subsubsection{Fine-tuning}
As depicted in Fig. 1(a), the output layer of the pre-trained FOP model is replaced by the combination of a global average pooling (GAP) layer and an emotion-specific projection layer. Then the whole model is fine-tuned for speech emotion recognition. Therefore, only parameters in the projection layer is trained from scratch in Fine-tuning.

We assume that a labeled sample consists of a mel-scale spectrogram $F$, along with an emotional label $y$. We first pass $F$ into the FOP model to obtain the outputs of the last attention layer $h_{L}$. Then $h_{L}$ is fed into the GAP layer and the projection layer with parameters $W_{y}$ to predict the emotional state $y$. This gives us the following objective to maximize:

\begin{equation}
	P(y|F)=softmax(GAP(h_{L})W_{y})
\end{equation}

\subsubsection{Hypercolumns}
As depicted in Fig. 1(b), the mel-scale spectrogram $F$ and the outputs of each attention layer $h_{l}, \; l \in [1,L]$ are utilized as different levels of abstraction for the waveform. As $F$ has the same feature dimensionality as $h_{l}, \; l \in [1,L]$, we further add and concatenate these representations together, which are marked as $h_{add}$ and $h_{concat}$, respectively.

\begin{equation}
h_{add}=F+\sum_{l=1}^Lh_l
\end{equation}
\begin{equation}
h_{concat}=Concat(F,h_1,...,h_L)
\end{equation}

\section{Databases}

\textbf{IEMOCAP}: The IEMOCAP \cite{busso2008iemocap} database contains about 12.46 hours of audiovisual data. There are five sessions with two actors each (one female and one male) and each session has different actors. Each session has been segmented into utterances, which are labeled into ten discrete labels (e.g., happy, sad, angry). To compared with other currently advanced approaches \cite{neumann2017attentive, lian2018speech}, we form a four-class emotion classification dataset containing $angry$, $happy$, $sad$ and $neutral$, where happy and excited categories are merged into a single happy category.

\textbf{VCTK}: The CSTR VCTK Corpus (Center for Speech Technology Voice Cloning Toolkit) \cite{yamagishi2012english} consists of 44 hours of data from 109 native speakers of English with various accents. Each speaker reads out about 400 sentences, which are selected from a newspaper plus the Rainbow Passage and an elicitation paragraph.

\section{Experiments and Results}

\subsection{Experimental setup}

In the experiments, we train a 2-layer FOP model with masked self-attention heads (256 dimensional states and 4 attention heads). For the feed forward layer in the FOP model, 513 dimensional inner states are chosen. We use the Adam \cite{kingma2014adam} optimization scheme with a learning rate of 0.001 and a batch size of 32. To prevent over-fitting, we use dropout \cite{srivastava2014dropout} with $p=0.2$ and early stopping \cite{prechelt1998automatic}. To alleviate the impact of the weight initialization, each configuration is tested 20 times. To consider the unbalanced number of samples between classes, weighted accuracy (WA) is chosen as our evaluation criterion.

\textbf{Source task (FOP)}: We utilize all samples (without labels) in the IEMOCAP and VCTK databases. 

\textbf{Target task (speech emotion recognition)}: We conduct experiments on the IEMOCAP dataset. There are five sessions and each session has different actors. To ensure that models are trained and tested on speaker independent sets, utterances from the first 2 speakers are used as the testing set and utterances from other speakers are used as the training set.

\subsection{Evaluation of FOP combined with Fine-tuning}
This section discusses the effectiveness of FOP combined with Fine-tuning. As the FOP model can be pre-trained on different datasets, three pre-training settings are discussed including ``VCTK'', ``IEMOCAP'' and ``None''. ``VCTK'' and ``IEMOCAP'' represent that the FOP model is pre-trained on the VCTK dataset and the IEMOCAP dataset, respectively. While the pre-training process is not adopted in ``None''.

\begin{table}[h]
	\centering
	\caption{Performance of FOP combined with Fine-tuning under different pre-training settings.}
	\label{table1}
	\begin{tabular}{cc}
		\hline
		Pre-training Settings 									& WA(\%)	          \\ \hline
		\textbf{None}: FOP without pre-training   				& 65.45           \\
		\textbf{VCTK}: FOP pre-trained on VCTK     				& 67.37           \\
		\textbf{IEMOCAP}: FOP pre-trained on IEMOCAP      		& \textbf{68.14}  \\ \hline
	\end{tabular}
\end{table}
Experimental results in Table 1 show that the pre-training process can improve the performance of speech emotion recognition. It reveals that after the pre-training process, the FOP model that learns to predict future information can also learn discriminative representations from waveforms. These high-level representations can be utilized to improve the recognition performance through Fine-tuning.
What's more, experimental results reveal that ``IEMOCAP'' gains higher classification accuracy than ``VCTK''. As speech emotion recognition is evaluated on the IEMOCAP dataset, the FOP model pre-trained on the IEMOCAP dataset can match with speech emotion recognition better than the FOP model pre-trained on the VCTK dataset. It indicates that, in transfer learning, the similarity of the source task and the target task can improve the performance of the target task.

Therefore, the FOP model is pre-trained on the unlabeled IEMOCAP dataset in the following experiments.

\subsection{Evaluation of FOP combined with Hypercolumns}
Hypercolumns extract the outputs of each attention layer in the FOP model, followed with additional classifiers for speech emotion recognition. In this section, we discuss the performance of FOP combined with Hypercolumns.

Multiple features (in Sec 2.2) are investigated. Since the mel-scale spectrogram $F$ is extracted without the FOP model, it is considered to be our baseline. In the 2-layer FOP model, the outputs of each attention layer are marked as $h_{1}$ and $h_{2}$, respectively. As $F$ has the same feature dimensionality as $h_{1}$ and $h_{2}$, we add and concatenate these representations together, which are marked as $h_{add}$ and $h_{concat}$, respectively. Besides various features, we also evaluate different classifiers for speech emotion recognition, including support vector machine (SVM), random forest (RF) and attention-based LSTM \cite{huang2016attention}.

\begin{table}[h]
	\centering
	\caption{Weighted accuracy (\%) of FOP combined with Hypercolumns under different combinations of features and classifiers. (Note: Baseline $F$ is extracted without the FOP model.)}
	\label{table2}
	\begin{tabular}{l|ccc|c}
		\hline
		Features	  & SVM 	& RF 	  & A-LSTM     & Best 				 \\ \hline
		$F$(Baseline) & 61.04   & 62.57   & 58.93      & 62.57 			 	 \\
		$h_{1}$       & 63.92 	& 56.66   & 64.68      & 64.68 			 	 \\
		$h_{2}$       & 63.53	& 57.39   & 62.38      & 63.53     			 \\
		$h_{add}$     & 63.72	& 58.35   & 64.88      & 64.88      		 \\
		$h_{concat}$  & 64.11	& 58.16   & 65.54      & \textbf{65.54}      \\ \hline
	\end{tabular}
\end{table}

Experimental results in Table 2 demonstrate that $F$ gains the worst performance, 62.57\%, among all features. It indicates that FOP can learn robust and generic representations (such as $h_{1}$, $h_{2}$) from the inputs $F$. These representations (such as $h_{1}$, $h_{2}$) are more effective than the inputs $F$ for speech emotion recognition. Therefore, we can conclude that transferring the ``knowledge'' from FOP into speech emotion recognition can improve recognition performance.

Furthermore, experimental results in Table 2 show that $h_{add}$ and $h_{concat}$ gain better performance than $h_{1}$, $h_{2}$ and $F$ in speech emotion recognition. $F$, $h_{1}$ and $h_{2}$ learn different levels of representations from the waveforms. As these representations are relevant to emotion recognition, $h_{add}$ and $h_{concat}$ that combine these emotion-related representations together can gain better performance.
What's more, we find that $h_{concat}$ is superior to $h_{add}$. The reason lies in that $h_{add}$ may result in the loss of original information in $h_{1}$, $h_{2}$ and $F$. While $h_{concat}$ not only can preserve original information, but also can combine these information properly.

\subsection{Comparison to other advanced approaches}
To show the effectiveness of the proposed method, we compare our method with currently advanced approaches through the five-folder cross validation.

\begin{table}[h]
	\centering
	\caption{Weighted accuracy (\%) of currently advanced approaches and the proposed approach on the IEMOCAP dataset.}
	\label{table3}
	\begin{tabular}{l|cc}
		\hline
													& Single-folder 	& Cross-validation \\ \hline
		Neumann et al. \cite{neumann2017attentive} 	& ---             	& 63.85            \\
		Lian et al. \cite{lian2018speech} 			& ---             	& 62.20            \\ 
		DAE in \cite{huang2018speech} 				& ---             	& 56.40            \\
		Ladder network in \cite{huang2018speech} 	& ---             	& 59.10            \\ 
		FOP+Hypercolumns							& 65.54        	 	& 63.56            \\
		FOP+Fine-tuning								& \textbf{68.14}    & \textbf{65.03}   \\ \hline
	\end{tabular}
\end{table}


Compared with our proposed method, these approaches \cite{neumann2017attentive, lian2018speech} also utilized mel-scale spectrograms as inputs, and showed promising results for speech emotion recognition. Neumann et al. \cite{neumann2017attentive} proposed an attentive CNN with multi-view learning objective function for speech emotion recognition. The CNN learned the representations of the audio signal, while the attention layer computed the weighted sum of all the information extracted from different parts. Lian et al. \cite{lian2018speech} introduced the contrastive loss function for speech emotion recognition. This loss function encouraged intra-class compactness and inter-class separability between different emotional categories. Experimental results in Table 3 demonstrate the effectiveness of the proposed method. Our method outperforms currently advanced approaches \cite{neumann2017attentive, lian2018speech} in weight accuracy. It proves that FOP, which learns discriminative representations from the original inputs, can be utilized to improve the performance of speech emotion recognition.

Meanwhile, we compare our method with other semi-supervised and unsupervised learning strategies. Huang et al. \cite{huang2018speech} utilized semi-supervised learning with ladder networks, which outperformed unsupervised DAE above 2\% for speech emotion recognition. Experimental results in Table 3 demonstrate that our method shows above 4\% performance improvement over the ladder network in \cite{huang2018speech}. DAE and ladder structures pay less attention to modeling long-term dynamic dependencies. However, temporal information is important for speech emotion recognition \cite{huang2016attention, kim2017towards}. Therefore, our self-attention based FOP model, which can capture long-term dynamic dependencies, is superior to other currently advanced unsupervised learning strategies.

\section{Conclusions}
This paper combines the unsupervised representation learning strategy -- FOP, with transfer learning approaches (such as Fine-tuning and Hypercolumns) for speech emotion recognition. To evaluate the effectiveness of the proposed method, we conduct experiments on the IEMOCAP database.
Experimental results reveal that FOP can learn robust and discriminative representations from the mel-scale spectrograms. These representations are more effective than the original mel-scale spectrograms for speech emotion recognition. 
As our method can capture long-term dynamic dependencies, it also outperforms the mainstream semi-supervised and unsupervised learning strategies (such as DAE and ladder structures) above 4\% for speech emotion recognition.

Future investigations include a detailed analysis of input features for FOP. Besides mel-scale spectrograms, other audio features (e.g., Mel Frequency Cepstrum Coefficients (MFCCs), Linear Predictive Codings (LPCs)) should be evaluated. Additionally we aim to investigate and further improve the classification accuracy using other advanced fine-tuning approaches (e.g., discriminative fine-tuning in \cite{howard2018universal}). Finally, we will apply the proposed method to other audio classification tasks, such as key word spotting and speaker identification.

\section{Acknowledgements}
This work is supported by the National Key Research \& Development Plan of China (No.2017YFC0820602), the National Natural Science Foundation of China (NSFC) (No.61425017, No.61831022, No.61773379, No.61771472), and the Strategic Priority Research Program of Chinese Academy of Sciences (No.XDC02050100).

\bibliographystyle{IEEEtran}

\bibliography{mybib}

\end{document}